\documentclass[10pt,A4paper,conference]{IEEEtran}

\usepackage{amsmath}
\usepackage{epsfig}

\begin{document}

\title{The Secret Key--Private Key Capacity Region for Three Terminals}

\author{\authorblockN{Chunxuan Ye}
\authorblockA{Department of Electrical and Computer Engineering\\
and Institute for Systems Research\\
University of Maryland\\
College Park, MD 20742, USA \\
E-mail: cxye@eng.umd.edu}
\and
\authorblockN{Prakash Narayan}
\authorblockA{Department of Electrical and Computer Engineering\\
and Institute for Systems Research\\
University of Maryland\\
College Park, MD 20742, USA \\
E-mail: prakash@eng.umd.edu} 
 }

\maketitle

\begin{abstract}
We consider a model for secrecy generation, with three terminals, by means of public interterminal communication, and examine the problem of characterizing all the rates at which all three terminals can generate a ``secret key,'' and -- simultaneously -- two designated terminals can generate a ``private key'' which is effectively concealed from the remaining terminal; both keys are also concealed from an eavesdropper that observes the public communication. Inner and outer bounds for the ``secret key--private key capacity region'' are derived. Under a certain special condition, these bounds coincide to yield the (exact) secret key--private key capacity region.
\end{abstract}

\section{Introduction}
The problem of secret key generation by multiple terminals, based on their observations of distinct correlated sources followed by public communication among themselves, has been investigated by several authors (\cite{Mau93, AhlCsi93, BenBra95, MauWol97, MauWol99, MauWol00, CsiNar00, MauWol03a, MauWol03b, MauWol03c, RenWol03}, among others). It has been shown that these terminals can generate common randomness which is kept secret from an eavesdropper that is privy to the public interterminal communication, and sometimes also to a wiretapped source which is correlated with the previous sources.

In the wake of \cite{Mau93}, \cite{AhlCsi93}, models for secrecy generation by multiple terminals have been widely studied. Of particular interest to us is recent work in \cite{CsiNar04}, which considers a model consisting of an arbitrary number of terminals that respectively observe the distinct components of a discrete memoryless multiple source (DMMS) followed by unrestricted public communication among themselves; a subset of the terminals can also serve as ``helpers'' for the remaining terminals in generating secrecy. Three varieties of secrecy capacity -- the largest rate of secrecy generation -- are considered according to the extent of an eavesdropper's knowledge: {\it secret key, private key} and {\it wiretap secret key} capacity. A secret key (SK) generated by a set of ``user'' terminals with assistance -- in the form of additional correlated information -- from a set of helper terminals (e.g., centralized or trusted servers in a key establishment protocol), requires concealment from an eavesdropper with access to the public interterminal communication. A private key (PK) generated by the user terminals must be additionally protected from the assisting helper terminals. A wiretap secret key\footnote{The capacity problem associated with a wiretap secret key is not fully resolved even in the case of two user terminals, and we do not consider it here.} must satisfy the even more stringent requirement of being protected from a resourceful eavesdropper's access to a wiretapped correlated source. It should be mentioned that in all of the work mentioned above, the user terminals are required to devise {\it only a single key}, of any variety, to be used subsequently for secure encrypted communication.

There are, however, situations, arising for instance in ``group communication,'' in which {\it multiple keys must be simultaneously devised in a coordinated manner by different groups of terminals} (with possible overlaps of groups); such keys need protection from prespecified terminals as also from an eavesdropper. For instance, in group communication, different groups of terminals (with possible overlaps of groups) must generate different keys for encrypted communication within those groups. A key devised for a group must be concealed from terminals outside that group as well as from an eavesdropper. Such ``group-wide'' keys can be simultaneously devised in a coordinated manner by different groups of terminals. Separate keys for different groups are also needed when certain disabled terminals become unauthorized or unreliable so that the keys assigned to them, in effect, are compromised; to maintain security, the remaining authorized terminals must then switch to another set of keys which are concealed from the disabled terminals. In the interests of efficiency, all such keys must be devised at the outset of operations so as to avoid the need for a fresh key generation procedure after a disablement. 

In general, in a network with $m$ terminals, we could have one (common) secret key for all the terminals, and private keys for every proper subset of the $m$ terminals. These situations produce a rich vein of secrecy generation problems, the information-theoretic underpinnings of which are substantial enough for investigation already in the case of just three terminals. The first work on the simultaneous generation of multiple keys is \cite{YeNar04}, in which the problem of generating two PKs for two different groups of user terminals is investigated.

In this paper, we consider a simple model with three terminals and examine the problem of characterizing all the rates at which the following two types of keys can be generated {\it simultaneously}: (i) all the three terminals generate a SK, which is effectively concealed from an eavesdropper; and (ii) a designated pair of terminals generate a PK, which is effectively concealed from the remaining terminal as well as the eavesdropper. Suppose that terminals ${\cal X}$, ${\cal Y}$ and ${\cal Z}$ observe, respectively, the distinct components of a DMMS, i.e., independent and identically distributed (i.i.d.) repetitions of the generic random variables (rvs) $X$, $Y$, $Z$, respectively. The terminals are permitted unrestricted communication among themselves over a public channel, and all the transmissions are observed by all the terminals. An eavesdropper has access to this public communication too, but gathers no additional (wiretapped) side-information; also, the eavesdropper is passive, i.e., unable to corrupt the transmissions. Terminals ${\cal X}$, ${\cal Y}$ and ${\cal Z}$ generate a SK, which is concealed from the eavesdropper with access to the public communication among the terminals. Also, terminals ${\cal X}$ and ${\cal Y}$ generate a PK, with the possible help of terminal ${\cal Z}$, which is concealed from the helper terminal ${\cal Z}$ and from the eavesdropper. The set of all rate pairs at which such (SK, PK) pairs can be generated is called (SK, PK)-capacity region.

Our main technical results are inner and outer bounds for the (SK, PK)-capacity region. Under a special condition, these bounds coincide to yield the (exact) capacity region.


\section{Statement of Results}

Consider a DMMS with three components corresponding to generic rvs $X$, $Y$, $Z$, with finite alphabets ${\cal X}$, ${\cal Y}$, ${\cal Z}$. Let $X^n=(X_1,\cdots ,X_n)$, $Y^n=(Y_1,\cdots , Y_n)$, $Z^n=(Z_1,\cdots , Z_n)$ be $n$ i.i.d. repetitions of the rvs $X$, $Y$, $Z$. The terminals ${\cal X}$, ${\cal Y}$, ${\cal Z}$ \footnote{The use of the same symbol for a terminal as well as for the alphabet of its observations should not lead to any confusion.} respectively observe the components $X^n$, $Y^n$, $Z^n$ of the DMMS $(X^n, Y^n, Z^n)$, where $n$ denotes the observation length. The terminals can communicate with each other through broadcasts over a noiseless public channel, possibly interactively in many rounds. Following \cite{CsiNar04}, we assume, without loss of generality, that these transmissions occur in consecutive time slots in $r$ rounds; the communication is depicted by $3r$ rvs $F_1, \cdots ,F_{3r}$, where $F_t$ denotes the transmission in time slot $t$, $1\leq t\leq 3r$, by a terminal assigned an index $i=t\mod 3$, $1\leq i\leq 3$, with terminals ${\cal X}$, ${\cal Y}$, ${\cal Z}$ corresponding to indices 1, 2, 3, respectively. In general, $F_t$ is allowed to be any function, defined in terms of a mapping $f_t$, of the observations at the terminal with index $i$, $i=t\mod 3$, and of the previous transmissions $F_{[1,t-1]}=(F_1,\cdots, F_{t-1})$; thus, for instance, $F_1=f_1(X^n)$, $F_2=f_2(Y^n, F_1)$, $F_3=f_3(Z^n, F_{[1,2]})$, and so on. We do not permit any randomization at the terminals; in particular, $f_1, \cdots, f_{3r}$ are deterministic mappings. Let ${\bf F}=(F_1,\cdots, F_{3r})$ denote collectively all the transmissions in the $3r$ time slots. 

Given $\varepsilon>0$ and the rvs $U$, $V$, we say that $U$ is {\it $\varepsilon$-recoverable} from $V$ if $\Pr \{U\neq f(V)\}\leq \varepsilon$ for some function $f(V)$ of $V$ (cf. \cite{CsiNar04}).

The rvs $K_{\cal S}$, $K_{\cal P}$, which are functions of $(X^n,Y^n,Z^n)$, with finite ranges ${\cal K}_{\cal S}$ and ${\cal K}_{\cal P}$, respectively, represent an {\it $\varepsilon$-(SK, PK) pair}, where the SK is for all the terminals and the PK is for terminals ${\cal X}$, ${\cal Y}$ with privacy from terminal ${\cal Z}$, achievable with communication ${\bf F}$, if:

$\bullet$ $K_{\cal S}$ is $\varepsilon$-recoverable from each of $({\bf F}, X^n)$, $({\bf F}, Y^n)$, $({\bf F}, Z^n)$; 

$\bullet$ $K_{\cal P}$ is $\varepsilon$-recoverable from each of $({\bf F}, X^n)$, $({\bf F}, Y^n)$; 

$\bullet$ $K_{\cal S}$ satisfies the secrecy condition and the uniformity condition
\begin{equation}
\frac{1}{n}I(K_{\cal S}\wedge {\bf F})\leq \varepsilon;
\label{c1}
\end{equation}
\[
\frac{1}{n}H(K_{\cal S})\geq \frac{1}{n}\log |{\cal K_{S}}|- \varepsilon;
\]
and

$\bullet$ $K_{\cal P}$ satisfies the secrecy condition and the uniformity condition 
\begin{equation}
\frac{1}{n}I(K_{\cal P}\wedge {\bf F}, Z^n)\leq \varepsilon;
\label{c2}
\end{equation}
\[
\frac{1}{n}H(K_{\cal P})\geq \frac{1}{n}\log |{\cal K_{P}}|- \varepsilon.
\]
The conditions above thus mean that terminals ${\cal X}$, ${\cal Y}$ and ${\cal Z}$ generate a nearly uniformly distributed SK $K_{\cal S}$ which is concealed from an eavesdropper that observes the public communication ${\bf F}$. Simultaneously, {\it based on the same public communication}, terminals ${\cal X}$ and ${\cal Y}$ generate a PK $K_{\cal P}$ with the terminal ${\cal Z}$ acting as helper (e.g., a ``third-party'' in a key establishment protocol) by providing ${\cal X}$, ${\cal Y}$ with additional correlated information; this private key is nearly uniformly distributed, and is concealed from an eavesdropper that observes the public communication ${\bf F}$ as well as from the helper ${\cal Z}$ (hence, ``private''). Note that the previous conditions readily imply that $K_{\cal S}$ and $K_{\cal P}$ are ``nearly'' statistically independent. 

{\bf Definition 1}: A pair of nonnegative numbers ($R_{{\cal S}}$, $R_{{\cal P}}$) constitute an {\it achievable (SK, PK)-rate pair} if for every $\varepsilon>0$ and sufficiently large $n$, an $\varepsilon$-(SK, PK) pair $\left(K_{\cal S}, K_{\cal P}\right)$ is achievable with suitable communication (with the number of rounds possibly depending on $n$), such that $\frac{1}{n}H\left(K_{\cal S}\right)\geq R_{{\cal S}}-\varepsilon$, $\frac{1}{n}H\left(K_{\cal P}\right)\geq R_{{\cal P}}-\varepsilon$. The set of all achievable (SK, PK)-rate pairs is the {\it (SK, PK)-capacity region}, denoted by ${\cal C}_{SP}$. 

\noindent {\it Remarks:} 

1. Maurer \cite{Mau94} pointed out that the secrecy conditions (\ref{c1}) and (\ref{c2}) were inadequate for cryptographic purposes, and should be strengthened by omission of the factor $\frac{1}{n}$. While all our achievability results below are presented in the ``weak sense,'' they can be established in the stronger sense of \cite{Mau94} by using the techniques developed in \cite{MauWol00}.

2. The (SK, PK)-capacity region ${\cal C}_{SP}$ is a closed convex set. Closedness is obvious from the definition, while convexity follows from a time-sharing argument (cf. \cite{CsiKor81}).

3. If $K_{\cal P}$ is set equal to a constant in the definition above, i.e., only a (single) $\varepsilon$-SK is generated by terminals ${\cal X}$, ${\cal Y}$ and ${\cal Z}$, then the entropy rate of such a secret key is called an {\it achievable SK-rate}, and the largest achievable SK-rate is the {\it SK-capacity}. It is known \cite{CsiNar04} that the SK-capacity is equal to 
\begin{equation}
\min\left\{\begin{array}{l}
              I(X\wedge Y,Z), I(Y\wedge X,Z), I(Z\wedge X,Y), \\
              \frac{1}{2}\left[H(X)+H(Y)+H(Z)-H(X,Y,Z)\right] 
            \end{array}
    \right\}.
\label{e9}
\end{equation}

4. If $K_{\cal S}$ is set equal to a constant in the definition above, i.e., only a (single) $\varepsilon$-PK is generated by terminals ${\cal X}$ and ${\cal Y}$ with terminal ${\cal Z}$ serving as a helper terminal, then the entropy rate of such a private key is called an {\it achievable PK-rate}, and the largest achievable PK-rate is the {\it PK-capacity}. It is known (cf. \cite{AhlCsi93}, \cite{CsiNar00}) that the PK-capacity is equal to 
\begin{equation}
I(X\wedge Y|Z).
\label{e10}
\end{equation}

\noindent {\sl Example 1:} Let $X$ and $Y$ be independent rvs, each uniformly distributed on $\{0,1\}$. Let $Z=X\oplus Y$, where $\oplus$ denotes addition modulo 2.

\noindent It is easily seen from (\ref{e9}) that the SK-capacity for the terminals ${\cal X}$, ${\cal Y}$, ${\cal Z}$ equals $\frac{1}{2}$, and from (\ref{e10}) that the PK-capacity for the terminals ${\cal X}$, ${\cal Y}$, with privacy from ${\cal Z}$, equals 1. We claim in this elementary example that 1 bit of {\it perfect} SK (i.e., $\varepsilon$-SK with $\varepsilon=0$) is achievable for all the terminals, with observation length $n=2$, using the following scheme. Terminals ${\cal X}$, ${\cal Y}$, ${\cal Z}$, with respective observations $(X_1,X_2)$, $(Y_1,Y_2)$, $(Z_1, Z_2)$, transmit $X_1$, $Y_2$ and $Z_1\oplus Z_2$, respectively. Then each terminal can perfectly recover all the observations of the other terminals. The secret key $K_{\cal S}$ is set to be $X_2$ (or $Y_1$ or $Z_1$ or $Z_2$). It can be shown that
\[
I(K_{\cal S}\wedge {\bf F})= I(X_2\wedge X_1, Y_2, Z_1\oplus Z_2)=0,
\]
and 
\[
H(K_{\cal S})=1.
\]
On the other hand, 1 bit of perfect PK is achievable for terminals ${\cal X}$ and ${\cal Y}$, with privacy from ${\cal Z}$, for observation length $n=1$. When terminal ${\cal Z}$ transmits ${\bf F}=Z_1$, terminal ${\cal X}$ can perfectly recover $Y_1$, which is set to be $K_{\cal P}$. It is clear that 
\[
I(K_{\cal P}\wedge {\bf F}, Z^n)=0,
\] 
and 
\[
H(K_{\cal P})=1.
\] 
Using a time-sharing argument, every (SK, PK)-rate pair $(R_{\cal S}, R_{\cal P})$ satisfying
\begin{equation}
2R_{\cal S}+R_{\cal P}\leq 1
\label{e11}
\end{equation}
is perfectly achievable. The results in this paper (cf. Theorem 1 below) will show that the secret key-private key capacity region ${\cal C}_{SP}$ for this example cannot be larger than the region in (\ref{e11}), so that (\ref{e11}) characterizes the capacity region ${\cal C}_{SP}$.  

\vspace{0.15in}

For notational simplicity, we set
\[
A\stackrel{\triangle}{=} I(Z\wedge X,Y),
\]
\[
B\stackrel{\triangle}{=} \min\left\{I(X\wedge Y,Z),I(Y\wedge X,Z) \right\},
\] 
\[
C\stackrel{\triangle}{=} \frac{1}{2}[H(X)+H(Y)+H(Z)-H(X,Y,Z)],
\] 
Thus, the SK-capacity (\ref{e9}) for the terminals ${\cal X}$, ${\cal Y}$, ${\cal Z}$ is equal to $\min\{A,B,C\}$.

{\bf Theorem 1} ({\it Outer bound for ${\cal C}_{SP}$}): Let $(R_{\cal S}, R_{\cal P})$ be an achievable (SK, PK)-rate pair. Then 
\begin{equation}
R_{\cal S}\leq A,
\label{thm1.1}
\end{equation}
\begin{equation}
R_{\cal P}\leq I(X\wedge Y|Z),
\label{thm1.2}
\end{equation}
\begin{equation}
R_{\cal S}+R_{\cal P}\leq B,
\label{thm1.3}
\end{equation}
\begin{equation}
2R_{\cal S}+R_{\cal P}\leq 2C.
\label{thm1.4}
\end{equation}

\noindent {\it Remark:} The bounds (\ref{thm1.1}), (\ref{thm1.2}) on the individual largest achievable SK- and PK-rates are obvious from (\ref{e9}) and (\ref{e10}). Also, while (\ref{e9}) implies 
\[
R_{\cal S}\leq B, \ \ \ \ \ R_{\cal S}\leq C,
\]
note that the conditions (\ref{thm1.3}), (\ref{thm1.4}) above are more stringent than (\ref{e9}).

{\bf Theorem 2} ({\it Inner bound for ${\cal C}_{SP}$}): The (SK, PK)-capacity region ${\cal C}_{SP}$ is inner-bounded by the region
\begin{equation}
  \left\{\begin{array}{ll}
    (R_{\cal S}, R_{\cal P}): & \frac{\min\{A,B,C\}-\min\{I(X\wedge Z), I(Y\wedge Z)\}}{I(X\wedge Y|Z)}\cdot R_{\cal P}\\
                              &+ R_{\cal S} \leq \min\{A,B,C\}, \\
                              & R_{\cal P}\leq I(X\wedge Y|Z)
                 \end{array}
          \right\}.
\label{thm1.5}
\end{equation}

\noindent {\it Remark:} The proof of Theorem 2 is based on the following idea: a modified version of the random binning technique developed in \cite{CsiNar04} is first used to generate the needed ``common randomness.'' A SK and a PK, of rate pair $\left(\min \left\{I(X\wedge Z), I(Y\wedge Z) \right\}, I(X\wedge Y|Z)\right)$ are then extracted from this common randomness, by a means from \cite{CsiNar04}. An application of the time-sharing technique then leads to the achievability of the region in (\ref{thm1.5}). Although interterminal communication between ${\cal X}$, ${\cal Y}$, ${\cal Z}$ is permitted, the region in (\ref{thm1.5}) is shown to be achieved by a single autonomous transmission from each terminal based on its own local observation of its component of the DMMS.

Under a certain condition, the outer bound in Theorem 1 coincides with the inner bound in Theorem 2, which provides a characterization of the (SK, PK)-capacity region ${\cal C}_{SP}$.

{\bf Theorem 3}: If $\min\{A,B,C\}=B$, then ${\cal C}_{SP}$ equals the set of pairs $(R_{\cal S}, R_{\cal P})$ satisfying (\ref{thm1.2}) and (\ref{thm1.3}).

\vspace{0.1in}

\noindent {\sl Example 2:} Let $X$, $Y$ and $Z$ be three rvs, each uniformly distributed on $\{0,1\}$, and satisfying the Markov condition $Y -\!\!\circ\!\!- X-\!\!\circ\!\!- Z$. Further, suppose that
\[
P_{XY}(x,y)=\frac{1-p}{2}\delta_{x,y}+\frac{p}{2}(1-\delta_{x,y}),
\]
\[
P_{XZ}(x,z)=\frac{1-q}{2}\delta_{x,z}+\frac{q}{2}(1-\delta_{x,z}),
\]
where $0<q<p<\frac{1}{2}$ and
\[
   \delta_{x,y}= \left\{\begin{array}{ll}
                 0,     &{\rm if}\ x\neq y, \\
                 1,    &{\rm if}\ x=y.
                 \end{array}
          \right.
\] 
Straightforward calculations show that
\[
A=I(Z\wedge X,Y)=1-h(q),
\]
\[
B=\min\{I(X\wedge Y, Z), I(Y\wedge X,Z)\}=1-h(p),
\]
and
\[
C=\frac{1}{2}[H(X)+H(Y)+H(Z)-H(X,Y,Z)]=1-\frac{h(p)+h(q)}{2},
\]
where $h(p)=-p\log_2p-(1-p)\log_2(1-p)$ is the binary entropy function. Since $0<q<p<\frac{1}{2}$, we have that $\min\{A, B,C\}=B$. It follows from Theorem 3 that ${\cal C}_{SP}$ is the set of pairs $(R_{\cal S}, R_{\cal P})$ satisfying
\begin{eqnarray*}
R_{\cal P}&\leq& h(p+q-2pq)-h(p),\\
R_{\cal S}+R_{\cal P}&\leq & 1-h(p).
\end{eqnarray*}
This region is depicted in Fig. 1.   
         \begin{figure}
           \epsfxsize=3in
           \centerline{\epsffile {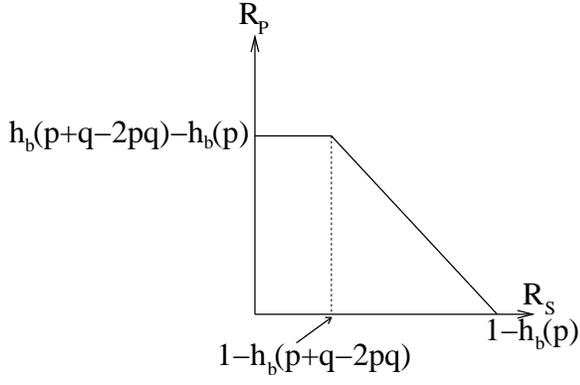}}
           \caption{${\cal C}_{SP}$ for Example 2.}
           \label{fig0}
         \end{figure} 

\noindent {\it Remarks:} Although we have shown the tightness of the outer bound for ${\cal C}_{SP}$ under the condition $\min\{A,B,C\}=B$, it remains open as to whether this outer bound is tight in general. To prove its tightness, it would suffice to show the tightness of the outer bound under the condition $\min\{A,B,C\}=\min\{A,C\}$. Since the case $A<B<C$ can be easily seen to be impossible, two remaining cases are relevant, and these are unresolved to date.

\noindent Case 1: $\min\{A, B, C\}=C$: Under this condition, the constraint (\ref{thm1.1}) is implied by the constraint (\ref{thm1.4}). Thus, the outer bound for the (SK, PK)-capacity region is given by the constraints (\ref{thm1.2}), (\ref{thm1.3}) and (\ref{thm1.4}), and is depicted in Fig. 2. By a time-sharing argument, to show the achievability of this region, it suffices to show that (SK, PK)-rate pairs $(0, I(X\wedge Y|Z))$, $(C,0)$, 
\[
\left(\min \left\{I(X\wedge Z),I(Y\wedge Z)\right\},I(X\wedge Y|Z)\right),
\]
and
\begin{equation}
(\max\{I(X\wedge Z), I(Y\wedge Z)\}, B-\max\{I(X\wedge Z), I(Y\wedge Z)\})
\label{thm1.13}
\end{equation}
are all achievable. While the first three (SK, PK)-rate pairs are known to be achievable, it is unclear if the achievability of (SK, PK)-rate pair (\ref{thm1.13}) holds.

         \begin{figure}
           \epsfxsize=3.5in
           \centerline{\epsffile {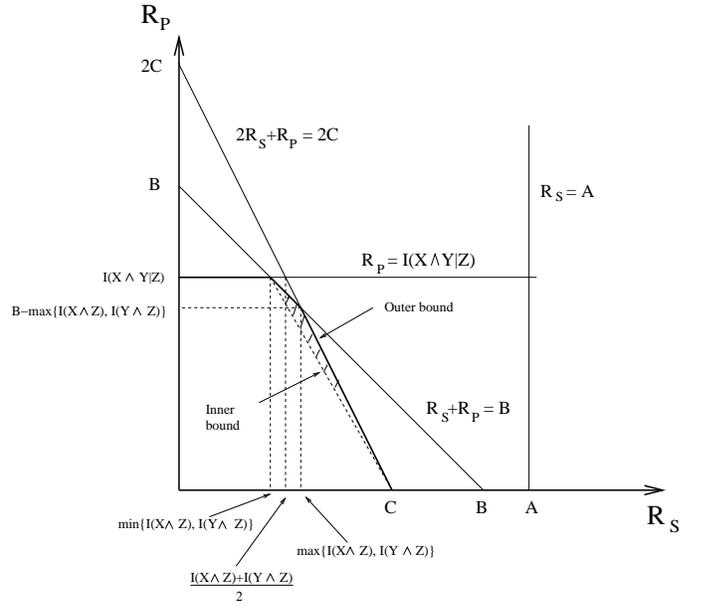}}
           \caption{Inner and outer bounds for ${\cal C}_{SP}$ for Case 1.}
           \label{fig2}
         \end{figure} 

\noindent Case 2: $A<C\leq B$: The outer bound for the (SK, PK)-capacity region under this condition is depicted in Fig. 3. To show that this region is achievable, it suffices to show the achievability of (SK, PK)-rate pairs $(0, I(X\wedge Y|Z))$, $(A,0)$, 
\[
\left(\min \left\{I(X\wedge Z),I(Y\wedge Z)\right\},I(X\wedge Y|Z)\right),
\]
\begin{equation}
(\max\{I(X\wedge Z), I(Y\wedge Z)\}, B-\max\{I(X\wedge Z), I(Y\wedge Z)\}),
\label{thm1.13a}
\end{equation}
and
\begin{equation}
\left(I(Z\wedge X,Y), I(X\wedge Y)-I(Z\wedge X,Y)\right).
\label{thm1.17}
\end{equation}
While the achievability of the first three (SK, PK)-rate pairs can be shown, it remains unclear whether (SK, PK)-rate pairs (\ref{thm1.13a}) and (\ref{thm1.17}) are achievable.

         \begin{figure}
           \epsfxsize=3.5in
           \centerline{\epsffile {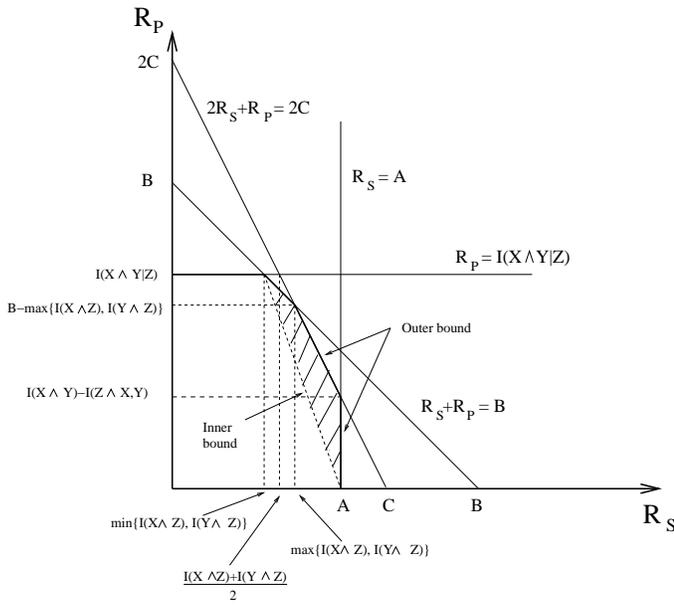}}
           \caption{Inner and outer bounds for ${\cal C}_{SP}$ for Case 2.}
           \label{fig3}
         \end{figure}



\section{Discussion}

Inner and outer bounds are derived for the (SK, PK)-capacity region for a model for secrecy generation with three terminals, each of which observes a distinct component of a discrete memoryless multiple source, with unrestricted public communication allowed among these terminals. Under a certain condition, these bounds coincide to yield the (SK, PK)-capacity region.

An obvious generalization of our model above is one in which a secret key is generated by all three terminals, and -- simultaneously -- all three pairs of terminals generate distinct private keys, each of which is effectively concealed from the remaining terminal. Entropy rates of these simultaneously generated secret key and private keys constitute a (SK, 3-PK)-rate quadruple. The set of all achievable (SK, 3-PK)-rate quadruples is called (SK, 3-PK)-capacity region. Following arguments similar to those used in the proof of Theorem 1, we can also obtain an outer bound for this (SK, 3-PK)-capacity region. Achievability proofs leading to inner bounds for this (SK, 3-PK)-capacity region are under investigation.

\IEEEtriggeratref{4}

\end{document}